# Constrained non-collinear magnetism in disordered Fe and Fe-Cr alloys

D. Nguyen-Manh[*], Pui-Wai Ma, M.Yu. Lavrentiev, S.L. Dudarev

*EURATOM/CCFE Fusion Association, Culham Science Centre, Abingdon, Oxon, OX14 3BD, United Kingdom*
[*] Corresponding Author, E-mail: duc.nguyen@ccfe.ac.uk

The development of quantitative models for radiation damage effects in iron, iron alloys and steels, particularly for the high temperature properties of the alloys, requires understanding of magnetic interactions, which control the phase stability of ferritic-martensitic, ferritic, and austenitic steels. In this work, disordered magnetic configurations of pure iron and Fe-Cr alloys are investigated using Density Functional Theory (DFT) formalism, in the form of constrained non-collinear magnetic calculations, with the objective of creating a database of atomic magnetic moments and forces acting between the atoms. From a given disordered atomic configuration of either pure Fe or Fe-Cr alloy, a penalty contribution to the usual spin-polarized DFT total energy has been calculated by constraining the magnitude and direction of magnetic moments. An extensive database of non-collinear magnetic moment and force components for various atomic configurations has been generated and used for interpolating the spatially-dependent magnetic interaction parameters, for applications in large-scale spin-lattice dynamics and magnetic Monte-Carlo simulations.

## I. Introduction

High-temperature properties of structural fusion materials, including those characterizing the response of materials to irradiation at high temperature, represent the key unknown entities critical to the development of viable fusion reactor design, and are a source of major uncertainty in the choice of reactor design strategy. Presently, significant experimental and modeling effort is devoted to understanding, testing and interpreting the available data on high-temperature properties of fusion materials, and to interpolating the data to conditions being not accessible to experiments and tests.

Iron-based alloys and steels, including ferritic-martensitic, ferritic, and austenitic steels, owe their phase stability to magnetic interactions. This fact is confirmed by DFT calculations performed in combination with both Cluster Expansion (CE) and Magnetic Cluster Expansion (MCE) methods[1-6]. Given the extensive use of steels in ITER and DEMO, and the requirement that these materials are expected to operate at temperatures exceeding 350°C, and possibly approaching 750°C, there is a genuine need to develop accurate quantitative understanding of radiation damage effects in iron, iron alloys and steels specifically in the high temperature limit. An additional complication associated with the treatment of magnetic effects in relation to radiation damage phenomena is that the initial development of radiation damage occurs through high-energy events, called collision cascades, where the energy of a primary recoil atom is transferred, through interatomic interactions, to the local environment of the recoil atom and exciting electrons, resulting in the local melting of the lattice, shock wave events, and the formation of vacancy and self-interstitial atom defects, as well as defect clusters. The challenge associated with modeling a cascade event in a magnetic material stems from the fact that not only the temperature of the material in a cascade is high, but also that the local atomic structure at any given moment of time no longer resembles that of a regular body-centered cubic (or face-centered cubic in the case of austenitic steels) lattice. To correctly understand the dynamics of evolution of a cascade, it is necessary to understand in detail the process of energy transfer between electronic, magnetic and atomic degrees of freedom in a quantitative form[7], including both the conventional "scalar" interatomic interactions and directional inter-magnetic interactions (i.e. exchange interactions between magnetic moments) for an atomic configuration disordered by a high-energy collision event. The objective of this work is to develop an accurate DFT database for simulating high-energy cascade events in Fe-based alloys by simultaneously taking into account both magnetic and atomic degrees of freedom. It also shows non-collinear magnetic excitations induces extra forces comparing to collinear case.

## II. Constrained Spin-Vector DFT method

To access non-collinear ground states properties of complex magnetic systems as well as for the prediction of finite-temperature properties from first-principles, vector-spin density functional theory (DFT) has to be applied, which treats the magnetization density as a vector field (and not as a scalar field, as in collinear DFT calculations). In 1972, von Barth and Hedin extended this concept to a spin-polarized system[8], replacing the scalar density by a Hermitian 2 x 2 matrix **n(r)**. From such calculations it is possible to follow several directions. Like in molecular-dynamic simulations, spin-dynamics allows to study the magnetic degrees of freedom either exploring the ground state or excited state properties. Or a model Hamiltonian approach is used where magnetic interactions are studied using parameters obtained from *ab initio* calculations. In both cases we introduce a discretization of the vector magnetization density: in spin-dynamics, the evolution of discrete spin vectors attached to certain atomic positions is monitored. Mapping the first-principles results onto a model Hamiltonian, which contains interactions between spins, also requires that it is possible to assign a definite spin to an atom. In the vicinity of an atom, e.g. within a sphere centered to the nuclei, it should be therefore possible to define the magnetization density

*M(r)=M$_i$e$_i$* (1)

where *M$_i$* is the magnitude of magnetization and *e$_i$* is the magnetization direction.

For the many-body Stoner model[3] and the Heisenberg-Landau MCE Hamiltonian model[4] we formulated a theorem, closely related to the concept of dynamic temperature of magnetic moments[9], stating that for an arbitrary non-collinear magnetic configurations with arbitrary exchange fields, at energy extremum each magnetic moment, *M$_i$*, associated with atom site i, is parallel to its effective exchange field, *H$_i$*, acting on it, i.e.,

*M$_i$*×*H$_i$*=0 (2)

For a specific (constrained) direction of magnetic field, {*H$_{i,c}$*}, the non-stationary solution of the total energy functional, E[n,{*M$_{i,c}$*}] is given by,

*E[n,{M$_{i,c}$*}]=*E$_{DFT}$* + *E$_p$({M$_i$},{M$_{i,c}$})* (3)

Here, the first term is the usual DFT total energy and the second term represents energy penalty associated with the locally constrained magnetic fields. {*M$_{i,c}$*} denotes the magnitude and direction of the desired (constrained) magnetic moments whereas {*M$_i$*} is an ensemble of integrated magnetic moments inside the sphere centered at the nucleus atom *i*. The constrained DFT developed by Dederichs *et al.*[10] provides the necessary framework to deal with arbitrary magnetic configurations, i.e. configurations where the orientations of the local moments are constrained to non-equilibrium directions. Non-constrained DFT calculations make it possible to find not only the spin-vector magnetic moments {*M$_i$*} but also provide additional information about inter-atomic forces

*F$_i$*=(δE[n,{*M$_{i,c}$*}]/δ*r$_i$*) (4)

acting on atoms under constrained magnetic fields.

## III. Application

The non-constrained DFT calculations were performed using the Vienna Ab-initio Simulation Package (VASP)[11,12] within the generalized gradient approximation (GGA) with the Perdew-Burke-Emzerhof exchange and correlation functional[13]. For our calculations of non-collinear magnetism, we used the spin-interpolation proposed by Vosko, Wilk and Nusair[14]. Solution of the Kohn-Shame equations have been carried out using a plane-wave basis set with an energy cut-off of 400 eV and with the projector augmented wave (PAW) pseudo-potentials in which semi-core electrons have been included. It is important to stress that, like for all the other bcc transition metals, the inclusion of semi-core electrons through the use of pseudo-potentials is important for predicting accurately the defect formation energies [15] in iron and iron-chromium based alloys.

From Eq. (3), the penalty term in the total energy, E[n,{*M$_{i,c}$*}], is given by

*E$_p$({M$_i$},{M$_{i,c}$})*=∑$_i$ λ[*M$_i$* - *M$_{i,c}$*(*M$_{i,c}$*.*M$_i$*)]$^2$ (5)

for constraining the direction of magnetic moments. The sum in Eq. (5) is taken over all atomic sites *i*, *M$_{i,c}$* is the desired direction of the magnetic moment at site *i*, and *M$_i$* is the integrated magnetic moment inside a sphere W$_i$ around the position of atom *i*. The penalty term introduces an additional potential inside the sphere of each atomic site. This potential is determined by a function of Pauli spin-matrices σ=(σ$_x$,σ$_y$,σ$_z$), {*M$_{i,c}$*}, {*M$_i$*} and the weight λ of the penalty energy. In the present study, a penalty functional is added to the system which derives the integrated local moments from the z-component of magnetization (0 0 1) into the specific direction (0 1 1).

### 1. Non-collinear magnetic configurations of disordered iron

The initial disordered atomic samples used in the present investigation were generated by molecular dynamic simulations for magnetic iron obtained from quenching, using conjugate gradient, of atomic structures at very high temperature (~10000K) for three different atomic densities corresponding to the bcc lattice constants of 2.75A°, 2.86A° and 3.00A°, respectively. In these simulations, the atoms interact via scalar many-body forces calculated from a magnetic interatomic potential[16] as well as via spin orientation dependent forces of the Heisenberg form. Spin-lattice dynamics simulations[9] showing the complexity of finite-temperature magnetic structure of a partially disordered ensemble of iron atoms, have now made it possible to explore the dynamics of magnetic moments for atomic configurations resembling those occurring in a hot liquid metal. However, the understanding of how magnetic moments interact in an atomically disordered structure of iron similar to that encountered in a collision cascade, is still qualitative and rudimentary. Fig. 1 shows the non-collinear magnetic configurations obtained from the constrained spin-vector DFT calculations for disordered iron configuration with 250 atoms for the three considered atomic densities. About 100 disordered configurations of pure Fe have been investigated and explored in the present work.

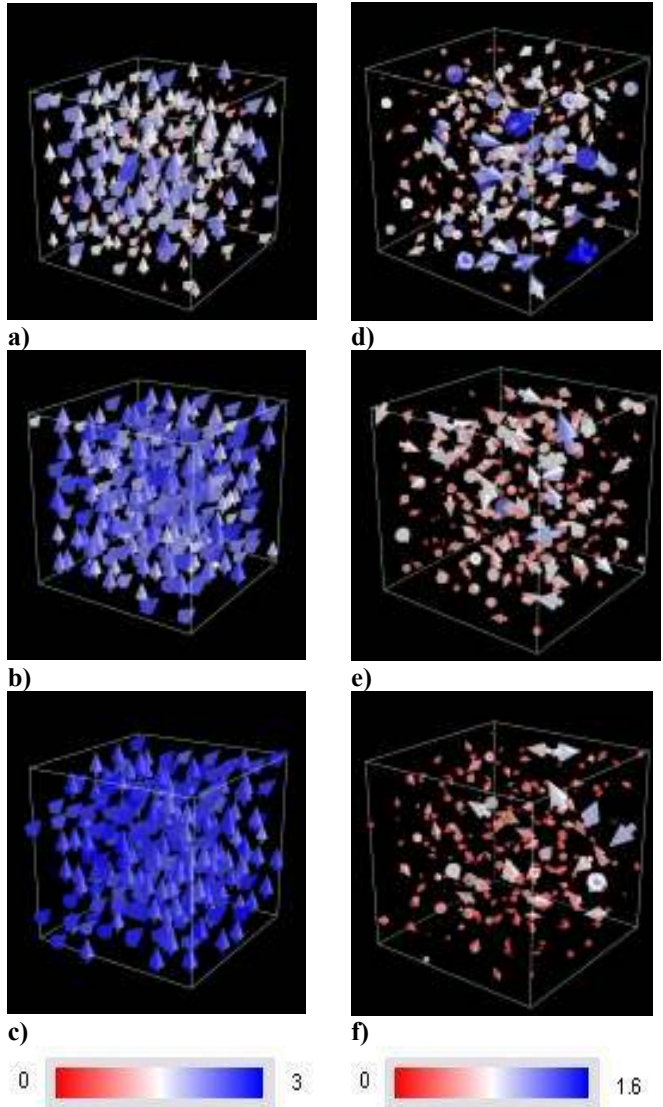

a)
b)
c)
d)
e)
f)

**Figure 1**: **Constrained non-collinear magnetic configurations in amorphous Fe obtained from DFT calculations in combination with MD simulations for 250 atoms at three different atomic densities with bcc lattice parameters: a) 2.75Å, b) 2.86Å and c) 3.00Å. The corresponding residual forces are shown in d), e) and f), respectively. The magnitude of magnetic moments and residual forces are represented by the respective colours in legends and by the lengths of arrows in figures.**

The constrained non-collinear magnetic configurations shown in Fig. 1 were obtained from self-consistent electronic structure calculations within the spin-vector DFT formalism but at fixed atomic positions generated from molecular-dynamic simulations. It is found that for the high atomic-density configurations ($a_{bcc}$=2.75Å), non-collinear magnetic moments are strongly frustrated (Fig. 1a) with few of them flipping into anti-ferromagnetic configurations with negative magnetization components. The corresponding residual force-vectors (Fig. 1d) are unusually large at several atomic sites reaching the maximum value of 1.6 eV/Å. The non-collinear magnetic configuration calculated at the equilibrium lattice constant for bcc iron ($a_{bcc}$=2.86Å) shows that all the magnetic moments are oriented along the desired positive direction (0 1 1) (Fig. 1b) and the majority of residual force-vectors (Fig. 1e) are smaller in magnitude in comparison with the high atomic density case. Finally, for the case of low atomic density ($a_{bcc}$=3.00Å), the constrained non-collinear magnetic moments are mainly positive with the maximum magnitude of 3.0 $\mu_B$ (Fig. 1c), whereas the residual force-vectors are mostly negligible except for very few atomic positions.

The above non-collinear configurations are being used as a new *ab initio* database of magnetic and atomic structures to develop an understanding required for interpolating the spatially-dependent magnetic interaction parameters between dissimilar structures, and applying them to large-scale spin-lattice dynamics simulations. In particular, this database serves for constructing and deriving more reliable functional form of inter-atomic potentials for iron with non-collinear magnetic moments based on the many-body Stoner tight-binding Hamiltonian[3,16,17].

### 2. Non-collinear magnetic configurations for Fe-Cr alloys

The magnetic origin of thermodynamic and kinetic phase transformations, including nano-clustering as well as the behaviour of point defects generated in irradiated Fe-Cr alloys have been systematically investigated by using a combination of DFT calculations with statistical approaches involving cluster expansions and Monte Carlo simulations[1,2,18]. Recently we have developed a new approach to the treatment of non-collinear magnetic order in Fe-based alloys, the Magnetic Cluster Expansion[4,5,19]. Using this approach we were able to predict several new effects, describing magnetic properties of nuclear steels, for example the composition and microstructure dependence of the Curie temperature, the non-collinearity of magnetic structures found in bcc Fe-Cr alloys, which has now been confirmed by experimental observations[20,21]. So far, the calculations forming the DFT database used for constructing the MCE model have been performed within the spin-polarized collinear magnetic framework at T=0K.

In this study, the constrained non-collinear magnetic DFT database is generated by using atomic configuration samples of binary Fe-Cr alloys obtained from exchange Monte-Carlo simulations with decreasing temperature as described in details previously and performed on small 4x4x4 bcc super-cells[3,18]. These configurations were found by cooling down, from 2000 K to 0 K, an initially random binary alloy configuration. The lowest negative mixing enthalpy is found at 6.25% Cr, which is consistent with our DFT prediction of an ordered Fe15Cr structure[1]. Importantly, despite considering only small supercell, it is found also that the mixing enthalpy changes sign as the Cr concentration increases and is positive everywhere above 10–12% Cr, where nano-clustering of the α' phase (bcc-Cr) is observed, in excellent agreement with both experimental and DFT data.

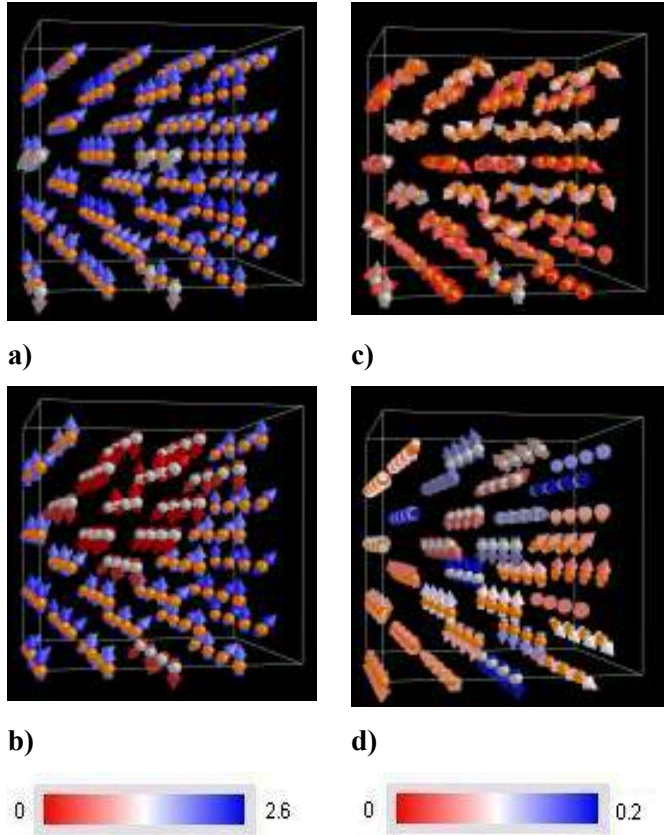

**Figure 2: Constrained non-collinear magnetic configurations in Fe-Cr alloys obtained from DFT calculations in combination with MC simulations for two different Cr compositions: a) 6.25% and b) 37.5% and their corresponding residual forces: c) and d), respectively**. The orange and grey colours denote Fe and Cr atom, respectively. Magnitudes of magnetic moments and residual forces are shown by colours indicated in the legends.

Here we performed calculations for the constrained non-collinear magnetic configurations with two different Cr concentrations: 6.25% and 37.5% (Fig. 2). For the case with low concentration, all the Cr atoms have their moments aligned anti-ferromagnetically in comparison with the magnetic moments associated with all the Fe sites, although the constraint of local magnetic field was imposed into (0 0 1) and ( 0 1 1) directions (Fig. 2a). The magnitude of the constrained non-collinear magnetic moments for Cr atoms varies from 1.3-1.6 $\mu_B$ in comparison with 1.7 $\mu_B$ for self-consistent collinear magnetic calculations. The resulting residual force-vectors are relatively small (Fig. 2c) with a maximum value of 0.2 eV/A° showing that the considered configuration of Fe-Cr alloys is not far from equilibrium condition.

At high concentration (37.5%Cr), nano-clusters of Cr atoms are formed due to the positive enthalpy of mixing related to the segregation of α' phase. It is shown that the Cr cluster is bounded by the equivalent (110) planes interfacing with the bcc Fe matrix. The lowest enthalpy of mixing configuration is thermodynamically consistent with the high density of (110) Cr/Fe interfaces. More importantly, as it has been analyzed earlier, spin-polarized DFT investigation demonstrates that the clustering phenomena in Fe-Cr binary with high Cr concentration have magnetic origin. Fig. 2c shows that the constrained fields on each atom imply the orientation of magnetic moments of Cr atoms aligning non-collinearly parallel to the (011) plane and anti-ferromagnetically orientated in comparison with those of Fe atoms. The magnitude of magnetic moments on Cr atoms is small (red colour) in comparison with high magnetic moments of 2.6 $\mu B$ on Fe atoms (blue colour). Despite complex and frustrated magnetic configurations at Cr/Fe interfaces, the residual forces found for the 37.5%Cr case (Fig. 2d) appear broadly similar to those evaluated for the case of 6.25%Cr from our constrained non-collinear DFT calculations.

We note that our present results of non-collinear magnetisms at interfaces in iron-chromium alloys are alternatives to those investigated previously by DFT calculations for the ground-state configurations[5,19]. For the (110) interfaces, the unconstrained DFT study showed that orientations of magnetic moments of Cr atoms are also aligned almost parallel to the interfaces but those of Fe atoms are oriented perpendicular to the interface. The constrained DFT non-collinear magnetic calculations shown in Fig. 2b and 2d provided additional information about the magnitude and orientation of magnetic moments as well as about the corresponding components of force vectors acting on each atom from the equilibrium and ground-state configurations. This information is important for the first-principles understanding of displacement per atom (dpa) behaviour in iron-based alloys and steels under irradiation conditions or cascade simulations, also offers an opportunity to extend the MCE formalism beyond the rigid-lattice approximation.

### 3. Non-collinear magnetic configurations for self-interstitial atom (SIA) defects

It is now well established that magnetic effects are also responsible for the fact that the atomic structure of radiation defects in iron and steels is different from the structure of defects formed under irradiation in non-magnetic body-centred cubic metals, for example vanadium or tungsten[6,15,17,22-24]. Unlike vacancies, SIAs do not spontaneously form in materials at elevated temperatures. SIAs generated in high-energy collision cascades migrate to sinks in the material, and this gives rise to swelling, irradiation creep and radiation embrittlement. Our DFT calculations of SIA defects in bcc transition metals show that in all the non-magnetic bcc transition metals, including bcc-W, the most stable defect configuration has the <111>

(crowdion) orientation. The pattern of ordering of SIA configurations is fundamentally different in ferromagnetic bcc-Fe, where the <110> dumbbell configuration is found to have the lowest formation energy. Spin-polarized collinear DFT calculations show that the two Fe atoms in the SIA <110> configuration have anti-ferromagnetically aligned magnetic moments of -0.25$\mu_B$ in comparison with the positive ferromagnetically ordered moments of 1.7 μB for the four nearest-neighbour iron atoms. Three-dimensional SIA clusters in iron formed directly in displacement cascades can be grown by capturing <110> dumbbells[25].

Fig.3a shows the result of DFT calculations within non-collinear magnetism, constraining the direction of magnetic moments for the <110> dumbbell defect in bcc-Fe. It is found that that the two iron atoms in a dumbbell configuration maintain the opposite orientation (red arrows) in comparison with the four nearest neighbour (grey arrows) ones although the direction of magnetic moments is different due to the local constrained magnetic field. The corresponding residual forces (Fig.3c) are almost negligible for two dumbbell atoms as well as for the surrounding atoms. We note that the initial configuration in this study was not the fully relaxed one obtained from collinear magnetic calculations without ionic relaxation.

If we have two Cr atoms in the <110> dumbbell configuration in bcc-Fe, their interaction with iron atoms results in a stabilized configuration in which the magnetic moments are larger and negative in comparison with those of the two iron atoms in the dumbbell within a collinear magnetic scheme[24]. Fig.3c shows the corresponding result within the constrained non-collinear magnetic configuration, where Cr atoms are represented by grey colour in comparison with the orange one for the Fe atoms. The orientation of magnetic moments on Cr atoms is again opposite to those of Fe atoms and the magnitude of these moments is about 0.7 $\mu_B$ compared with the maximum value of 2.6 $\mu_B$ for Fe atoms. Fig. 3d shows the corresponding residual force-vectors for Cr-Cr dumbbell defect in bcc-Fe. It is interesting to note that within the constrained non-collinear magnetic calculations the two Cr atoms in a dumbbell attract, with forces acting between the atoms pointing towards each other, whereas the four surrounding Fe atoms have the force direction pulling them away from the SIA defect configuration. Such artificial magnetic excitations induce extra forces comparing to collinear case, which is an concrete evidence that show the important of considering thermal excitation in magnetic subsystem, when we are calculating interatomic forces in magnetic metals.

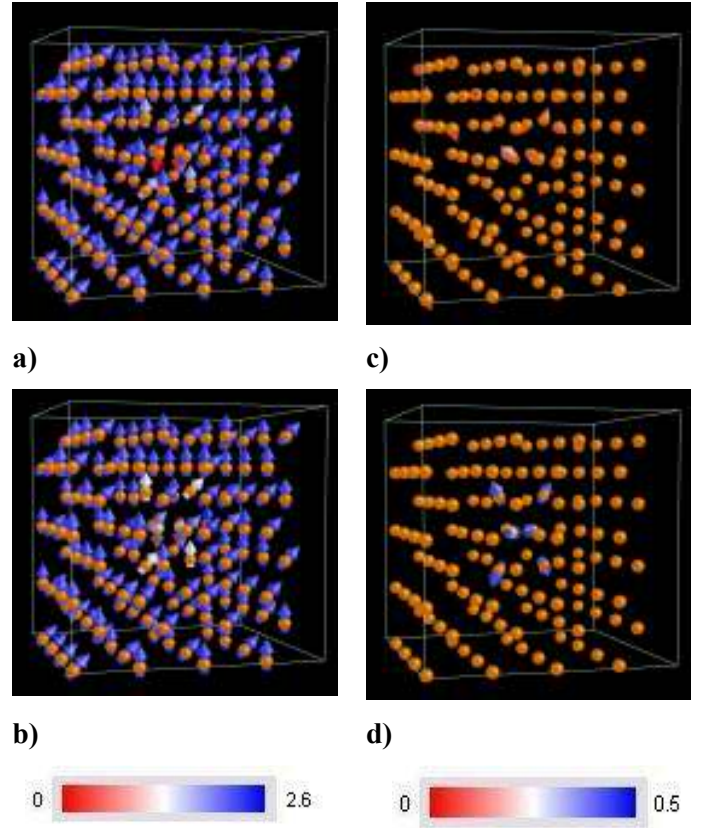

a)   c)

b)   d)

**Figure 3**: Constrained non-collinear magnetic configuration for <110> self-interstitial atom (SIA) defect configuration obtained from DFT calculations in bcc-Fe: a) Fe-Fe dumbbell, b) Cr-Cr dumbbell and their corresponding residual forces: c) and d), respectively. The orange and grey colours denote Fe and Cr atom, respectively. The magnitude for the magnetic moments and residual forces is shown by colours in legends.

## IV. Conclusion

We have performed extensive constrained spin-vector DFT calculations for disordered Fe, Fe-Cr alloy as well as for irradiation-induced SIA configurations for point defects with Fe-Fe and Cr-Cr <110> dumbbells, for the purpose of establishing a database of magnetic and atomic properties required for the parameterization of spatially-dependent magnetic interactions, with the subsequent application of the result to large-scale spin-lattice dynamics simulations of collision cascades. In addition to the DFT database containing ground-state configurations evaluated within collinear magnetic approximation, the constrained non-collinear calculations provide a much broader DFT data base containing very rich information about magnetic moments and forces (both treated as vectors with variable magnitude and direction in three-dimensional space) for disordered and defect configurations obtained under various non-equilibrium conditions. The present database is being used to develop many-body interatomic potentials for iron and iron-alloys with non-collinear treatment of magnetic moments in these systems. We also show forces can be induced by magnetic excitations.

**Acknowledgment**

This work, part-funded by the European Communities under the contract of Association between EURATOM and CCFE, was carried out within the framework of the European Fusion Development Agreement. To obtain further information on the data and models underlying this paper please contact *PublicationManager@ccfe.ac.uk*. The views and opinions expressed herein do not necessarily reflect those of the European Commission. This work was also part-funded by the RCUK Energy Programme under grant EP/I501045. DNM would like to thank the Juelich supercomputer centre for providing access to High-Performances Computer for Fusion (HPC-FF) facilities as well as the International Fusion Energy Research Centre (IFERC) for using the supercomputer (Helios) at Computational Simulation Centre (CSC) in Rokkasho (Japan).